\documentclass[letter]{aa}
\usepackage{epsfig}
\usepackage{graphics}
\usepackage{subfigure}
\usepackage{natbib}
\usepackage{amsmath}
\usepackage{amssymb}
\usepackage{upgreek}
\usepackage{color}
\usepackage{enumerate}
\usepackage{hyperref}
\hypersetup{
    colorlinks=true,
    linkcolor=blue,
    urlcolor=blue,
    citecolor=blue
}

\newcommand{\mJy}{\mathrm{mJy}}

\begin{document}

\title{Towards the origin of the radio emission in AR~Sco, the first radio-pulsing white dwarf binary}

\author{B.~Marcote\inst{1}, T.~R.~Marsh\inst{2}, E.~R.~Stanway\inst{2}, Z.~Paragi\inst{1}, J.~M.~Blanchard\inst{1}}
  
\institute{
Joint Institute for VLBI ERIC, Postbus 2, 7990 AA Dwingeloo, The Netherlands
\and
Department of Physics, University of Warwick, Gibbet Hill Road, Coventry CV4 7AL, UK
}

\offprints{B.~Marcote\\ \email{marcote@jive.eu}}

\titlerunning{AR~Sco on milliarsecond angular scales}

\authorrunning{Marcote et al.}

\abstract
{The binary system AR~Sco contains an M star and the only known 
radio-pulsing white dwarf. The system shows emission from radio to X-rays, likely dominated by synchrotron radiation. The mechanism that produces most of this emission remains unclear. Two competing scenarios have been proposed: Collimated outflows, and direct interaction between the magnetospheres of the white dwarf and the M star.}
{The two proposed scenarios can be tested via very long baseline interferometric radio observations.}
{We conducted a radio observation with the Australian Long Baseline Array (LBA) on 20 Oct 2016 at 8.5~GHz to study the compactness of the radio emission. Simultaneous data with the Australian Telescope Compact Array (ATCA) were also recorded for a direct comparison of the obtained flux densities.}
{AR~Sco shows radio emission compact on milliarcsecond angular scales ($\lesssim 0.02~\mathrm{AU}$, or $4\ \mathrm{R_{\odot}}$). The emission is orbitally modulated, with an average flux density of $\approx 6.5~\mathrm{mJy}$. A comparison with the simultaneous ATCA data shows that no flux is resolved out on mas scales, implying that the radio emission is produced in this compact region. Additionally, the obtained radio light curves on hour timescales are consistent with the optical light curve.}
{The radio emission in AR~Sco is likely produced in the magnetosphere of the M star or the white dwarf, and we see no evidence for a radio outflow or collimated jets significantly contributing to the radio emission.}

\keywords{binaries: close  -- white dwarfs -- Radiation mechanisms: non-thermal -- Radio continuum: general -- Techniques: high angular resolution -- Techniques: interferometric} 
 
\maketitle

\section{Introduction}

Jet-like outflows are a ubiquitous feature of accreting systems from the supermassive black holes at the center of galaxies to the stellar-mass black holes and neutron stars in close binary star systems, and including young stars in star forming regions. Accreting white dwarfs however seem by and large to be exceptions to the rule. Despite being common enough to be nearby and accessible to detailed study, for the most part accreting white dwarfs show little evidence for highly collimated outflows, although there have been suggestions of directed outflows in some high-accretion-rate systems \citep{2004MNRAS.347..430B}. Jets from black holes and neutron stars are associated with radio emission, and so the detection of radio emission has been used as indirect evidence for jet outflows from some accreting white dwarfs \citep{2008Sci...320.1318K,2011MNRAS.418L.129K}, however such a direct connection has yet to be proven. In the vast majority of cases, cataclysmic variable stars are not strong radio sources. A survey of 121 magnetic systems using the Karl G. Jansky Very Large Array (VLA) detected only 19 of them, all but one having a flux density below $1\,\mJy$ \citep{2017arXiv170207631B}. A study of disk accreting systems in their high states detected several, but with flux densities $< 0.1\,\mJy$ \citep{2016MNRAS.463.2229C}. The one exception, the system AE~Aqr, which can exceed $10\,\mJy$, is believed to be in an unusual ``propellering'' state in which its rapidly spinning magnetic white dwarf is flinging mass transferred from its companion out of the binary \citep{wynn1997}, and it is thought that this expelled gas is the source of its radio emission \citep{meintjes2012}.

Very recently the first rival to AE~Aqr in terms of its persistent radio flux was discovered \citep{marsh2016}. The system, AR~Scorpii (AR~Sco) is a close binary system composed of an M star and a white dwarf with an orbital period of 3.56~h. 
The system hosts the first and only radio pulsing white dwarf discovered so far. Its emission pulses periodically on a period of 1.97~min, thought to reflect the spin period of the white dwarf \citep{marsh2016}.
AR~Sco, which is located at a distance of $116 \pm 12~\mathrm{pc}$, exhibits emission from radio to X-rays, with a broadband spectral energy distribution (SED) characteristic of synchrotron emission.
The optical and radio emission are both modulated by the 1.97~min spin period, and the optical emission is also strongly modulated on the 3.56-h orbital period. The radio flux density of AR~Sco, which was observed to vary from 6 to $17\,\mJy$ at 9~GHz in an hour's run on the Australian Telescope Compact Array \citep[ATCA,][]{marsh2016}, is very similar to that of AE~Aqr, which at a similar frequency was observed to vary from 3 to 17~mJy during two six-hour-long runs on the VLA \citep{2012MmSAI..83..651M}. AR~Sco's orbital modulation features a maximum in optical flux around 0.1 orbital cycles before inferior conjunction of the white dwarf (when it is closest to Earth at orbital phase $\phi = 0.5$). Much of the emission appears to originate in the magnetosphere of the M star on the side facing the white dwarf \citep{marsh2016, katz2017}.
The ATCA data presented in the discovery paper were too short in duration to test whether AR~Sco's radio emission is also affected by the binary orbit.
The observed emission (with a luminosity $\langle L \rangle \approx 1.7 \times 10^{25}\ \mathrm{W}$) is predominantly from the synchrotron component which is powered by the spin-down luminosity of the white dwarf \citep{marsh2016}. The X-ray luminosity is only $\sim 4\%$ of its optical luminosity, which together with the lack of Doppler-broadened optical emission lines indicates that there is little or no accretion in the system. In this respect it differs significantly from AE~Aqr, which shows erratic flaring behavior in overall flux and line emission that is rooted in mass transfer and ejection.

The pulses in AR~Sco are observed from radio to ultraviolet wavelengths, showing a spin-down of $P \dot P^{-1} \sim 10^7\ \mathrm{yr}$, while the cooling time of the white dwarf is $\gtrsim 10^9~\mathrm{yr}$ \citep{marsh2016,beskrovnaya2016}. Strong optical linear polarization (up to $40\%$) and small but significant circular polarization have been observed, both being modulated by the orbital and spin periods \citep{buckley2017}. The presence of this polarization suggests a global topology for the magnetic field in the emission region, and may be indicative of a rapidly rotating white dwarf with a strong magnetic field, as seen in neutron star pulsars such as the Crab.

Most of the emission appears to arise on or near the M star, however it is not clear how the energy is transferred from the white dwarf to the M star. Two possibilities have been proposed so far \citep{marsh2016,buckley2017,katz2017}: Collimated relativistic particle outflows, and direct interaction between the white dwarf magnetosphere and the M star. Some form of collimation or specificity is required because the observed synchrotron emission amounts to $\sim 7$\% of the total spin down power, whereas the M star only covers $\sim 2$\% of the sky as seen from the white dwarf, ruling out isotropic ``irradiation'' as the transfer mechanism \citep{marsh2016}. This, along with the synchrotron spectrum, raises the question of whether the radio emission in AR~Sco comes from a relativistic jet, interesting given the often-used association between radio emission and jet outflows referred to earlier.

In this letter, we report on a very long baseline interferometric radio observation with the Australian Long Baseline Array (LBA) that allowed us to study the compactness of the radio emission from AR~Sco. We present the observations and the data reduction in Sect.~\ref{sec:obs}. We quote the results of these observations in Sect.~\ref{sec:results} and discuss the properties of the observed radio emission and the constraints that these results place on the origin of the radio emission in Sect.~\ref{sec:discussion}. Finally, we state our conclusions in Sect.~\ref{sec:conclusions}.

\begin{figure*}
    \begin{center}
        \includegraphics[width=0.76\textwidth]{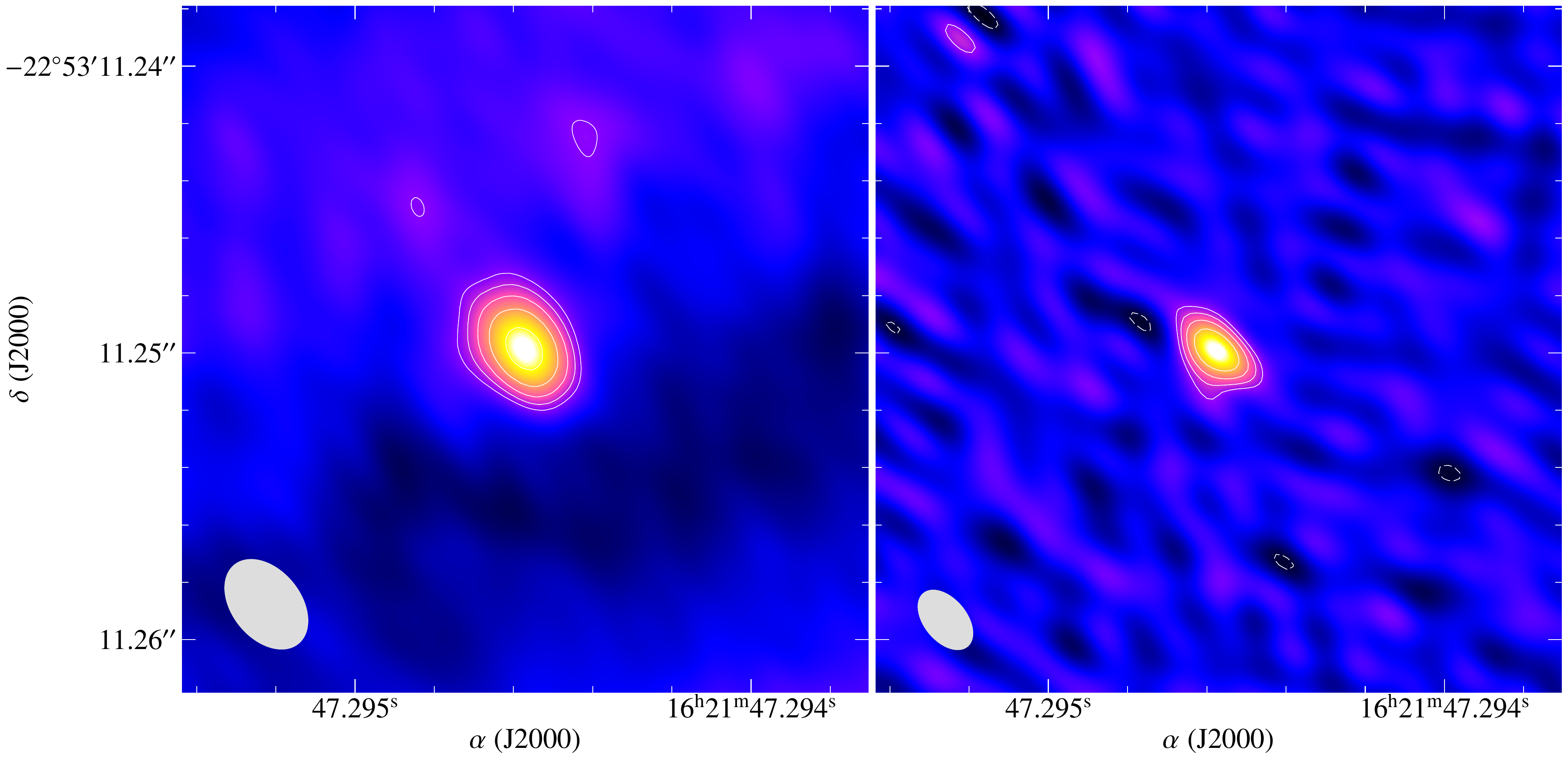}
        \caption{Images of AR~Sco obtained with the LBA on 20 October 2016 at 8.4~GHz. {\em Left:} Image obtained with a natural weighting and no self-calibration. {\em Right:} Image obtained with a Briggs robust weighting of zero after self-calibration in phase. Contours start at three times the noise level of 0.4~mJy in both cases and increase by factors of $2^{1/2}$. The synthesized beams, represented by the white ellipses in the bottom left corners, are $1.4 \times 2.6\ \mathrm{mas^2},\ {\rm PA} = 39^{\circ}$, and $1.8 \times 2.9\ \mathrm{mas^2},\ {\rm PA} = 40^{\circ}$, respectively.}
        \label{fig:lba}
    \end{center}
\end{figure*}

\section{Observations and data reduction}\label{sec:obs}

We observed AR~Sco at 8.4~GHz on 20 October 2016 from 0:20 to 9:00 UTC with the LBA (project code V548A). Seven stations participated during this observation: ATCA, Ceduna, Hobart, Mopra, Parkes, Katherine, and Warkworth. The data were recorded with a total bandwidth of 32~MHz divided into two subbands of 32 channels each, with full polarization. The ATCA also recorded simultaneous interferometric data with a larger bandwidth of 2~GHz to study the emission on arcsecond angular scales.
We used 3C~273  as a fringe finder, and J1625$-$2527 as the phase calibrator (located $2.7^{\circ}$ away from AR~Sco and with a flux density of $\sim 1.2\ \mathrm{Jy}$). A phase-referencing cycle of 5~min was used, expending 3.5~min on AR~Sco and 1.5~min on J1625$-$2527.

The LBA data have been reduced in AIPS\footnote{The Astronomical Image Processing System (AIPS) is a software package produced and maintained by the National Radio Astronomy Observatory (NRAO).} and Difmap \citep{shepherd1994} following standard procedures. A-priori amplitude calibration was performed using the known gain curves and system temperature measurements recorded on each station during the observation. For Warkworth we used the nominal System Equivalent Flux Density (SEFD).
Manual flagging was first performed to remove bad data (those mainly affected by Radio Frequency Interference). We then fringe-fitted and bandpass calibrated the data using 3C~273. We imaged and self-calibrated the phase calibrator, J1625$-$2527, to improve the final calibration of the data. Finally, the obtained solutions were transferred to AR~Sco. Different weighting schemes were used during the imaging. The final images were obtained by using a Briggs robust weighting of zero \citep{briggs1995}, which provided the optimal balance between resolution and sensitivity, and a natural weighting, which allowed us to self-calibrate the data in phase due to the higher sensitivity. This self-calibration improved the phase solutions, producing thus a more accurate image of the source.

We characterized the emission observed in the final images following two different approaches: by fitting a Gaussian component to the $uv$-data, which provides a more reliable measurement of the source size, and by fitting a Gaussian component in the image plane.
Since in very-long-baseline interferometric (VLBI) observations we cannot use primary flux calibrators (they are resolved on mas scales), our LBA flux density measurements reported below may be in error at the 10--20\% level.

The ATCA data have been reduced in Miriad \citep{sault1995} and CASA\footnote{The Common Astronomy Software Applications, CASA, is also produced and maintained by the NRAO.} following standard procedures. The amplitude
calibration of all stations was set using 3C~273. We then calibrated the data in phase using J1625$-$2527. The solutions in amplitude and phase were then transferred to AR~Sco, which was finally imaged.

\section{LBA and ATCA results}\label{sec:results}

As can be seen in Fig.~\ref{fig:lba}, AR~Sco is detected in the LBA data as a radio source compact on milliarcsecond angular scales.
Figure~\ref{fig:lba} (left) shows the image obtained with a natural weighting. AR~Sco is detected at the position $\alpha_{\rm J2000} = 16^{\rm h} 21^{\rm m} 47.294570(5)^{\rm s},\ \delta_{\rm J2000} = -22^{\circ} 53^\prime 11.24987(9)^{\prime\prime}$ with a flux density, measured in the image plane, of $6.5 \pm 0.7\ \mathrm{mJy}$. The rms noise level of the image is 0.4~mJy. A fit in the $uv$-data converges into a source with a flux density of $\approx 7.3\ \mathrm{mJy}$ and a size of 1.2~mas (compatible with the synthesized beam of $2.4 \times 3.5\ \mathrm{mas^2}$, position angle, PA, of $39^{\circ}$). 

We note the presence of phase errors in the data that can introduce artificial features in the source structure, a well-known caveat of phase-referencing observations (e.g., see the large-scale anti-symmetric amplitudes in Fig.~\ref{fig:lba}, left, around the source).
To improve the results we conducted a phase self-calibration of the data using 30-min solution intervals.
A natural weighting provided a more sensitive image with a point-like source of $6.8 \pm 0.2~\mathrm{mJy}$ (rms of 0.16~mJy). A Briggs robust weighting of zero provided a map with a higher resolution (see Fig.~\ref{fig:lba}, right) with a synthesized beam of $1.8 \times 2.9\ \mathrm{mas^2}, {\rm PA} = 40^{\circ}$. No extended emission is detected above the rms of 0.4~mJy.
A fit to the $uv$-data converges to a point like source of $\approx 7.1\ \mathrm{mJy}$ and a size of $\sim 0.17~\mathrm{mas}$ ($\sim 10\%$ of the synthesized beam). We note that the consistent solutions obtained across the sub-bands after self-calibration, the compatible flux densities obtained for AR~Sco before and after self-calibration, and the lack of any visible extended emission in the original data provide further confidence in the self-calibrated results.

The ATCA data present a much larger synthesized beam ($14 \times 13\ \mathrm{arcsec^2}, \mathrm{PA} = 83^{\circ}$), showing compact radio emission with a flux density of $6.2 \pm 0.2\ {\mathrm{mJy}}$, consistent with that obtained from the LBA data on mas scales.

\begin{figure}
    \begin{center}
        \includegraphics[width=0.46\textwidth]{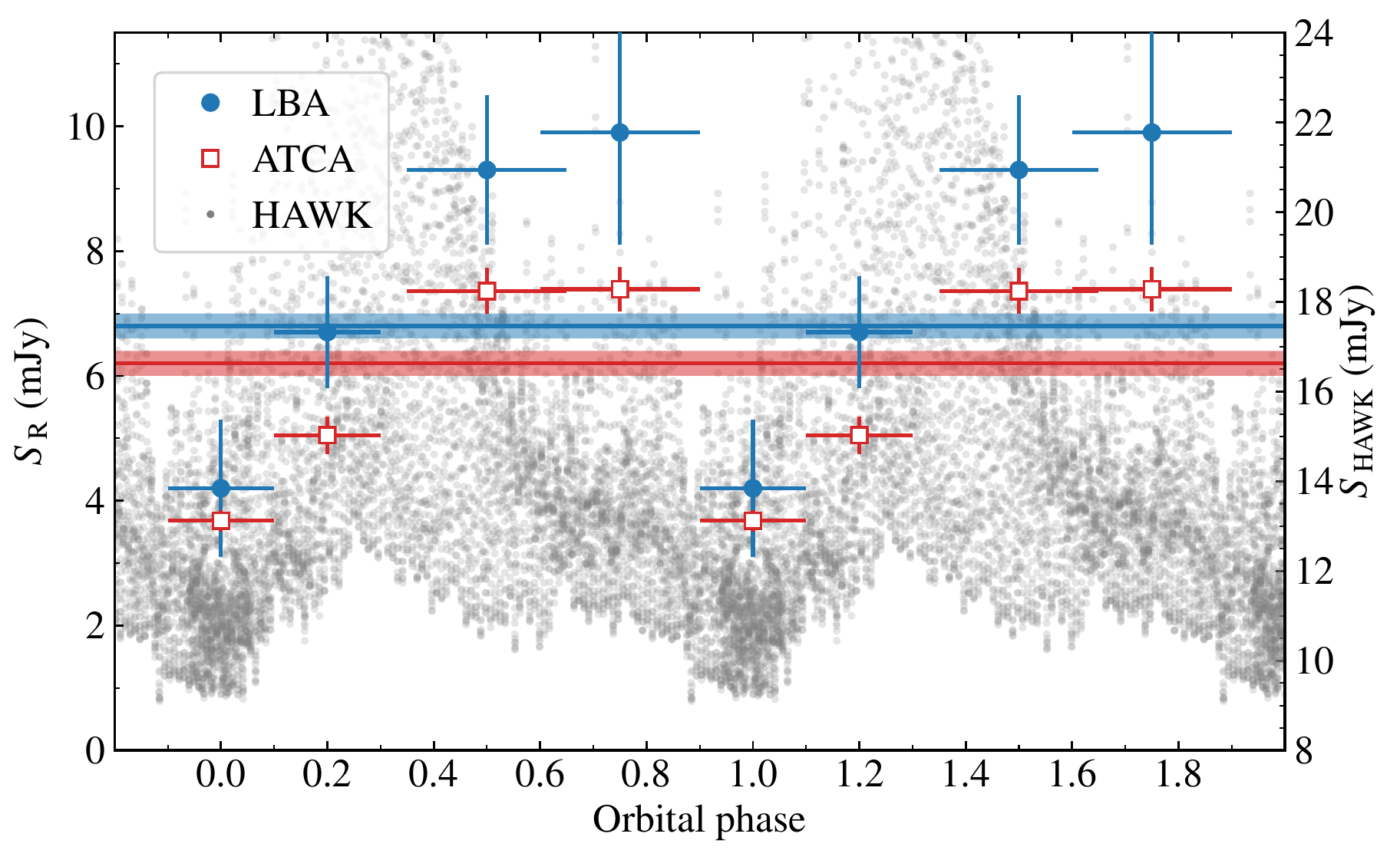}
        \caption{Light curve of AR~Sco derived from the LBA and ATCA data at 8.5~GHz and the VLT HAWK-I data at Ks band (2.14~$\mathrm{\upmu m}$) published by \citet{marsh2016}. The horizontal lines represent the average flux densities obtained from the full observation. Fluxes at specific orbital phases have been obtained by splitting the observation and combining different orbital cycles.}
        \label{fig:lc}
    \end{center}
\end{figure}
Almost 2.5 orbital cycles of AR~Sco were covered in our observation. This allows us to search for variability within the observation while keeping reasonable $uv$-coverage in both the LBA and the ATCA data sets. Figure~\ref{fig:lc} shows the light curve of AR~Sco as a function of the orbital phase during the performed observation as obtained from the LBA and ATCA data, using the ephemerides provided by \citet{marsh2016}. We observe a significant variability in the ATCA data ($> 6\sigma$). The LBA data show the same trend but with a significance of $2.8\sigma$ due to the larger uncertainties. The minimum of the radio emission takes place at an orbital phase of $\approx 0.0$, and is coincident with the minimum of the optical emission \citep[see Fig.~\ref{fig:lc} and][]{marsh2016}.

AR~Sco thus shows the same behavior along the orbital phase on both arcsec and mas scales, and all flux densities agree within 2$\sigma$. The LBA flux densities exhibit slightly but consistently higher values ($20$--$30\%$), which can be understood by a systematic offset applied between ATCA and LBA calibrations. Indeed, the obtained fluxes in the calibrator J1625$-$2525 exhibit a 30\% deviation between both datasets confirming this.

\section{Discussion}\label{sec:discussion}

\subsection{Astrometry of AR~Sco}

The position of AR~Sco obtained from the LBA data shows a statistical uncertainty of $\sim 0.01~\mathrm{mas}$. 
However, systematic errors must be taken into account for further comparisons. First, the uncertainty in the position of the phase calibrator, J1625$-$2527, is 0.2 and 0.1~mas in right ascension and declination, respectively \citep{beasley2002}. Secondly, the phase-referencing technique also introduces additional uncertainties to the final positions. Using the estimations provided by \citet{pradel2006} in phase-referencing VLBI observations, we expect an uncertainty of $\sim 0.3\ \mathrm{mas}$ in our image of AR~Sco.
As a conclusion, we would expect a total uncertainty of $\sim 0.5\ \mathrm{mas}$ on the measured position.

The position of AR~Sco reported in {\em Gaia} DR1 \citep{gaia2016} is $\alpha_{\rm J2015.0} = 16^{\rm h} 21^{\rm m} 47.293745^{\rm s} \pm 0.23\ {\rm mas},\ \delta_{\rm J2015.0} = -22^{\circ} 53^\prime 11.157505^{\prime\prime} \pm 0.45\ {\rm mas}$. This position was obtained from measurements over the interval 25 July 2014 and 16 September 2015 and is referenced to J2015.0, although, as AR~Sco is not bright enough to appear in the {\em Tycho} catalog, it does not have a full five-parameter astrometric solution in {\em Gaia} DR1. The uncertainties in the listed position are based upon modeling with a Bayesian prior on the proper motion and parallax \citep{2015A&A...583A..68M}.
The fourth U.S. Naval Observatory CCD astrograph catalog (UCAC4, \citealt{zacharias2013}) provides proper motion estimates of $\mu_{\alpha} = 12.4 \pm 3.0\ \mathrm{mas\ yr^{-1}},\ \mu_{\delta} = -50.5 \pm 2.4\ \mathrm{mas\ yr^{-1}}$ for AR~Sco. Correcting the radio position for parallax assuming a distance of $116\,$pc, we find that the LBA minus {\em Gaia} DR1 position difference is $\Delta\alpha = -4 \pm 5\ \mathrm{mas},\ \Delta\delta = -3 \pm 4\ \mathrm{mas}$ (without correcting for parallax 
the values are $\Delta\alpha = -10$ and $\Delta\delta = -1$). The uncertainties are dominated by the uncertainties in the ground-based proper motion measurements and can be expected to decrease significantly once a full five-parameter solution becomes available with {\em Gaia} DR2. At present the optical and radio positions match within their uncertainties.

\subsection{Compactness of the radio emission}

AR~Sco shows a compact radio emission on milliarcsecond scales with no signatures of extended emission. The model fit to the self-calibrated data provides a source size of 0.17~mas. Given the synthesized beam of the image ($1.8 \times 2.9\ \mathrm{mas^2}$), this value is comparable to the maximum resolution reachable by an interferometer \citep{martividal2012}. We thus conclude that the radio emission from AR~Sco is unresolved on mas scales.
This size implies that the radio emitting region is $\lesssim 0.02~\mathrm{AU}$ (or $4\ \mathrm{R_{\odot}}$, which is $\sim 3$ times the semimajor axis of the orbit). We thus infer a brightness temperature of $T_{b} \gtrsim 5 \times 10^9\ \mathrm{K}$, clearly implying non-thermal emission.

AR~Sco shows a compatible (at 2-$\sigma$ level) flux density of $\sim 6.5\ \mathrm{mJy}$ in both the LBA and the ATCA data, implying that there is no significant flux lost on mas scales. This supports the idea that most (if not all) the radio emission arises from a compact region near the white dwarf or at the surface of the M star \citep[as suggested by][]{marsh2016,katz2017}.

\subsection{Variability}

AR~Sco exhibits radio and optical light curves modulated by both the orbital and spin period \cite[see][in prep]{marsh2016,stanway2017}. The simultaneous LBA and ATCA data show that the radio emission on arcsecond and mas scales is synchronized in time on $\sim 20$-min timescales (see Fig.~\ref{fig:lc}). This adds further support to the lack of significant outflows or collimated jets in the system. If they were to exist we would expect those outflows to be detected in the LBA data, or to be resolved out. In the latter case we would expect a) lower emission on mas scales and b) a possible time delay between the two light curves (or no correlation at all) given the dynamical timescales of the particles producing the radio emission.

\section{Conclusions} \label{sec:conclusions}

We observed the radio-pulsing white dwarf binary AR~Sco with the Australian Long Baseline Array and Australian Telescope Compact Array, and found it to be unresolved on milliarcsecond scales. If the energy transport between the white dwarf and its companion M dwarf is via a jet, then either that jet must not contribute significantly to the electromagnetic flux from the system, or it must fade on a scale of a few solar radii. Alternatively, there is no jet, and the energy flow is sustained through magnetic interaction, perhaps inductive in nature, with most dissipation occurring close to the M dwarf.
Our observations contain evidence for variability of a factor of two in radio flux on the orbital cycle, with the time of minimum flux matching the phase of minimum optical flux.

\section*{Acknowledgments}

The Australian Long Baseline Array is part of the Australia Telescope National Facility which is funded by the Australian Government for operation as a National Facility managed by CSIRO.
BM acknowledges support from the Spanish Ministerio de Economía y Competitividad (MINECO) under grants AYA2016-76012-C3-1-P and MDM-2014-0369 of ICCUB
(Unidad de Excelencia “María de Maeztu”).
TRM and ERS acknowledge support from the Science and Technology Facilities Council (STFC) under grant ST/L000733.

\bibliographystyle{aa}
\bibliography{references}

\end{document}